# Constructing Evidence-Based Tailoring Variables for Adaptive Interventions

John J. Dziak, Ph.D. (0000-0003-0762-5495)

Institute for Health Research and Policy, University of Illinois Chicago, Illinois

dziak@umich.edu*

Inbal Nahum-Shani, Ph.D. (0000-0001-6138-9089)

Institute for Social Research, University of Michigan, Ann Arbor, Michigan

inbal@umich.edu

* Corresponding author. Authors contributed equally.

**Acknowledgments:** The authors thank Dr. David Vanness, Dr. Linda Collins, and Jamie Yap for helpful discussions, Ian Burnette for graphical design assistance, Dr. Caitlin Hayward for logistical guidance, and Amanda Applegate for careful editing suggestions. This work was supported by NIH awards P50 DA 054039 and NIH R01 DA 039901 from the National Institute on Drug Abuse. The content is solely the responsibility of the authors and does not necessarily represent the official views of the funding institutions. Part of the work was done while the corresponding author was working at the Institute for Health Research and Policy, University of Illinois Chicago.

**Authors' Statement of Conflict of Interest and Adherence to Ethical Standards**: John J. Dziak and Inbal Nahum-Shani state that they have no conflict of interest.

**Authors' Contributions**: John J. Dziak (Conceptualization [equal], Methodology [equal], Writing – original draft [equal], Writing – review & editing [equal], Visualization [equal]), Inbal Nahum-Shani (Conceptualization [equal], Methodology [equal], Writing – original draft [equal], Writing – review & editing [equal], Visualization [equal]).

**Keywords**: Experimental Design, Randomized Clinical Trials, Multiphase Optimization Strategy, Sequential Multiple Assignment Randomized Trials, Adaptive Interventions, Factorial Designs.

## Abstract

**Background**: Adaptive interventions provide a guide for using ongoing information about individuals to decide whether and how to modify the type, amount, delivery modality, or timing of treatment, to improve intervention effectiveness while reducing cost and burden. The variables that inform treatment modification decisions are called *tailoring variables*. Specifying a tailoring variable requires describing what should be measured, when to measure it, when the measure should be used to make decisions, and what cutoffs should be used in making decisions. These questions are causal and prescriptive (what to do, when), not merely predictive. They involve tradeoffs between specificity and sensitivity, and between waiting for sufficient information versus intervening quickly.

**Purpose**: There is little specific guidance in the literature on how to empirically choose tailoring variables, including cutoffs, measurement times, and decision times.

**Methods:** We review possible approaches for comparing potential tailoring variables and propose a framework for systematically developing tailoring variables.

**Results:** Although secondary observational data can be used to select tailoring variables, additional assumptions are needed. A specifically designed experiment for optimization (an optimization randomized controlled trial), e.g., a multi-arm randomized trial, sequential multiple assignment randomized trial, factorial experiment, or hybrid design, may provide a more direct way to answer these questions.

**Conclusions:** Using randomization directly to inform tailoring variables provides the most direct causal evidence but requires more effort and resources than secondary data analysis. More research is needed on how best to design tailoring variables for effective, scalable interventions.

## Introduction

To prevent and treat chronic conditions such as substance use [1], obesity [2], and mental health disorders [3, 4], it is critical to address the changing needs of individuals over time. This goal can be accomplished via adaptive interventions [5, 6]. Adaptive interventions are a family of intervention designs (i.e., procedures guiding intervention delivery in practice) that use ongoing information about the unit of interest (e.g., individual, dyad, group, organization) to decide whether and how to intervene. Formally, an adaptive intervention is a protocol composed of a sequence of decision rules that specify, for each decision time (i.e., each time at which intervention decisions should be made), which treatment to offer, when, for whom, and how. Adaptive interventions seek to improve effectiveness and resource efficiency by offering the appropriate type and amount of treatment when and only when it is needed.

Tailoring variables are key elements in adaptive interventions [5, 7]. A tailoring variable is information used to decide whether, when, or how to offer a treatment, or how much treatment to offer, to a specific person. The details of tailoring variables (timing, cutoff, etc.) may seem less tangible than the intervention components being offered to participants (e.g., coaching, messaging, education). However, given limited resources, even very high-quality interventions will not reach their potential if offered to those who would not benefit from them, rather than to those who need and can benefit from them. Thus, tailoring variables are integral components of ADIs, alongside the treatments whose delivery is being tailored [5, 7]. Tailoring variables can often be described in terms of four components. The first two focus on what to assess and when:

- **Observed Variable**(s): Which variable(s) should be measured for possible use in tailoring?
- **Assessment Time**(s): When should these variables be measured?

The second two focus on when and how assessments should be used to make a decision:

- **Decision Time**(s): When should the variables be used to make intervention decisions?
- **Cutoff(s)**: What threshold values on the observed variables best differentiate between those who should or should not be offered a given intervention option?

Investigators planning to develop effective, scalable adaptive interventions may have scientific questions about how best to define tailoring variables. Clinical experience and expertise, as well as a theoretical or conceptual model for the intended process of change, are vital in answering these questions, [5] but open scientific questions may remain. Further empirical information could be obtained either from secondary data analysis (SDA) or from optimization randomized controlled trials (ORCTs), a range of experimental designs for systematically assessing the performance of intervention components [8].

An ORCT may be a two- or multi-arm randomized controlled trial (RCT) [see 9], factorial experiment [8, 10, 11], sequential multiple assignment randomized trial (SMART) [7, 12-14], micro-randomized trial (MRT) [15], or hybrid design [16], depending on the adaptive intervention of interest and the scientific questions. ORCTs are often intended to inform the selection of components with high individual and combined performance, for use in an intervention package which could later be compared to a suitable control (e.g., standard of care) in an evaluation RCT [8, 10]. The components may include the treatments themselves but may also include the tailoring variables. Several considerations can be used to evaluate interventions; we focus here on effectiveness, but others may include cost and scalability [8].

In this paper, we provide a framework to guide the construction of tailoring variables. This framework delineates the respective advantages and drawbacks of each approach—SDA and ORCT—to inform tailoring variables, and underscores that choosing an approach should be guided by a conceptual model and clinical experience [7, 8]. While there are various types of adaptive interventions [17], we focus for simplicity on developing tailoring variables for classical adaptive interventions (abbreviated as ADIs) which address conditions that change relatively slowly (e.g., over a few weeks or months) and which include only a few decision points spaced relatively far apart in time.

Many variables could be used as tailoring variables for ADIs. For simplicity, we will focus on a common kind of tailoring variable: response status [5, 12]. This is a dichotomous indicator, assessed during treatment (after initiation but before completion), of whether the participant is showing sufficient ("responder") or insufficient early improvement ("nonresponder"). This is often defined by a cutoff (threshold) on some observed variable. The ADI might recommend that responders continue with initial treatment while nonresponders are offered an alternative, modified, augmented, or intensified treatment, informally called "rescue" treatment. Effectiveness calls for offering rescue to those most likely to benefit, while efficiency calls for avoiding rescue among those unlikely to benefit. Achieving these objectives depends on a clear, well-chosen definition of response status.

The operationalization of response status includes observed variable(s), assessment time(s) and decision time(s) (which may or may not coincide), and cutoff(s). Selecting these components can be complicated because the meaning of each depends on the others. For example, a variable must be chosen and measured before a cutoff can be applied. For clarity, we discuss the components in reverse order, starting with the cutoff, and we assume investigators have a question focused on one component, with all other components and treatment options already specified.

## Selecting a Cutoff

Consider an ADI for reducing cannabis use that starts with support from a mobile app and then uses information about app usage to decide whether to augment with human coaching.

Suppose app usage will be monitored continuously. A researcher may ask whether it is better to use a higher cutoff (accessing the app less than *twice* per week by week 4) or a lower cutoff (less than *once* per week by week 4) when determining who should be offered coaching. That is,

> **Question 1:** In an ADI starting with app alone, and then offering coaching to nonresponders starting at week 4, should participants be classified as nonresponders if they used the app less than $c = 1$ or less than $c = 2$ times per week by week 4, in order to obtain the best expected week-10 final outcome $Y$?

Notice that this is a causal question comparing the average potential outcomes (i.e., the population-level week-10 outcome under each alternative scenario) that *would be* expected *if* the researcher chooses one cutoff versus the other. SDA can provide some indirect information about this question, but an ORCT can address it more directly. We describe both options below.

**Comparing cutoffs using SDA**

Suppose the investigator has access to observational data on past individuals from the population of interest who were offered the initial treatment under consideration (e.g., app) for 10 weeks but were *not* offered the rescue treatment under consideration (e.g., coaching). The researcher might use this data to compare the average outcome for participants having less than, say, $c = 2$ app uses weekly, versus the average outcome for only those participants with less than $c = 1$ use weekly. However, this comparison (based on individuals not exposed to rescue treatment) may not indicate the optimal cutoff for a future ADI where that rescue treatment is employed, for two reasons. First, it cannot establish the needed balance of sensitivity versus specificity. A higher cutoff represents a more inclusive definition of nonresponse (more participants labeled as nonresponders); a lower cutoff represents a stricter definition (fewer nonresponders, perhaps with more severe needs). Selecting the optimal operational definition involves considering clinical priorities and resource constraints, not only empirical observational data. Second, predictive comparison does not address causality or potential outcomes. Comparing cutoffs by their ability to predict outcomes *without* rescue does not necessarily inform what would have happened *with* rescue, nor the implications of the cutoffs for rescue effectiveness.

A researcher could choose the cutoff associated with the highest in-sample prognostic accuracy (fewest false positives and false negatives) in predicting who is at high risk for an unsuccessful long-term outcome. However, this assumes that falsely concluding a person needs rescue is equally bad as falsely concluding they do not. Consider an extreme case where the rescue treatment is highly effective and has only minimal cost or burden. Here, it is reasonable to use a broad definition of nonresponse, prioritizing sensitivity, so that the treatment is made available to anyone who might benefit. In contrast, suppose the rescue treatment is slightly effective but highly costly and time-consuming and has burdensome side effects. This might suggest prioritizing specificity, offering rescue only when clearly necessary. This is a pragmatic tradeoff, involving potential consequences of different actions, not only a predictive question (see Table 1). Thus, the best cutoff for offering rescue treatment cannot be determined from secondary data under the initial treatment alone, without assumptions about the effectiveness of the rescue across different cutoffs, as well as the effort, cost, and burden associated with the rescue treatment.

Fewer assumptions may be needed if the SDA also includes some individuals who receive rescue treatment as well as some who do not. However, limitations remain. For example, suppose participants in the existing dataset were given rescue treatment at week 4, if and only if their app use was lower than once weekly at that time. Since the same cutoff was applied to all participants, there is no data about what would happen if the cutoff had been *twice* weekly instead, or at a different decision week. Those questions would require richer data, perhaps from a trial in which assignment to augmented treatment was random or a range of cutoffs was used. Existing methods for estimating the effectiveness of ADIs from observational data, as a form of causal inference [18-22], can estimate the average causal effect of rescue, conditional on different levels of the tailoring variable. However, using these methods to identify the tailoring variable values for which a rescue treatment is potentially beneficial requires several assumptions, including positivity (i.e., all treatment options were *a priori* possible for every participant). This implies that treatment probabilities—though potentially dependent on covariates—are above 0% and below 100% [18, 20, 21, 23]. This may not hold when comparing cutoffs if the dataset does not include one of the cutoffs of interest. Ideally, at least some participants were treated in accordance with each cutoff of interest [18], but finding such a dataset could be challenging if the rescue is a novel treatment. If the existing data are inadequate for their question, investigators may consider designing an appropriate trial to address it directly.

**Comparing cutoffs using ORCTs**

An ORCT to address Question 1 could initially offer all participants the app only, and at week 4, randomize them to either the lower ($c = 1$) or higher ($c = 2$) cutoff. Those who do not meet their assigned cutoff are considered nonresponders and are offered coaching; others continue with app only. The lower- and higher-cutoff arms could then be compared directly, to determine which cutoff leads to a higher average outcome. The average must be marginal (combined) over responders and nonresponders, not only nonresponders. Because each arm defines nonresponders differently, a comparison of "nonresponder" subgroups between them would be ambiguous.

## Selecting a Decision Time

Other common questions about tailoring involve choosing the time(s) at which tailoring decisions will be made. For simplicity in this section, assume the investigator is interested in designing an ADI with a single decision time, at which participants will be classified as responders or nonresponders according to a prespecified cutoff, and assume that rescue treatment will be offered to nonresponders soon after the decision. For the cannabis example, suppose that weekly app usage is continually assessed during an initial period and that individuals who use the app less than twice weekly on average will be considered nonresponders. Suppose the observed variable (weekly app usage) and the cutoff have already been specified, but the investigator is unsure about the decision time and therefore asks:

> ***Question 2*****:** Would the ADI be more effective if the decision is made at week 2 based on response status by week 2, or made at week 4 based on response status by week 4?

As before, either SDA or an ORCT might be used to address this question.

**Comparing decision times using SDA**

If the secondary data comes from a study in which participants receive only the initial treatment, then it cannot directly answer Question 2. It *can* be used to test whether weekly app usage at week 4, versus at week 2, is more *predictive* of the final outcome (see Table 1). More generally, an investigator could use longitudinal data to identify which time $t$ is characterized by the strongest correlation between an observed variable $O_t$ assessed at time $t$, and the final outcome $Y$ assessed at the end of the study at time $T_{\max}$. This might be described as finding the most predictive time $t$. However, predicting the outcome under the initial intervention alone does not determine whether or how the outcome would be modified by rescue treatment. The decision time that best predicts the final outcome may not be the best time for intervention delivery. Recall that for many health-related variables, measurements closer to each other in time are more highly correlated than those further apart. Thus, the predictive relationship between the intermediate observation and the final outcome will likely strengthen as $t$ approaches $T_{\max}$, especially if $O_t$ represents the same underlying concept as $Y$ or a closely related concept. The relationship strength might level off (perhaps $O_t$ no longer changes much after a certain time, or perhaps the correlation attenuates due to attrition-related bias or noise) but would be unlikely to meaningfully weaken. Thus, all else being equal, individuals' weekly cannabis usage at, say, week $T_{\max} = 10$ are probably more strongly predicted by behavior in week $t = 4$ (6 weeks lag) than in week $t = 2$ (8 weeks lag). Investigators might conclude from this that the rescue decision should not be made until the more informative week 4; but perhaps waiting until week 4 would lead to participant disengagement and undermine the benefits of the rescue treatment. This suggests a need to balance predictive accuracy (which might be improved by delaying the decision) with effective action (which might be improved by acting promptly). Focusing only on prognostic performance could lead to the selection of decision times too late for maximally effective change.

Heuristically, an "elbow plot" of predictive strength, across possible decision times, could be informative here. The goal would *not* be to find the time which gives the most accurate predictions, but instead the earliest time when a "good enough" predictive value (a "point of diminishing returns") is reached. Ideally, there will be a clear "elbow," before which the curve is sharply increasing and after which it levels off. This would be an intuitive choice for the decision time. Measures of predictive accuracy for use in an elbow plot include odds ratios (useful if a cutoff has already been chosen for each decision point) or the area under the receiver operating curve (ROC AUC) if cutoffs are also to be determined [24]. Unfortunately, there is no consensus criterion for exactly defining the elbow. More importantly, the elbow plot still does not provide direct information about the implications of different decision times on the effectiveness of rescue.

A conceptual model for how, when, and how quickly the rescue treatment is believed to work could help provide clarity. Richer secondary data in which some participants access the rescue treatment at different times might allow a causal analysis, under sufficient assumptions about confounding [21]. However, without a detailed conceptual model or sufficiently rich data, a separate investigation may be needed. An experiment in which decision times are randomly assigned could allow direct empirical estimation of causal effects.

**Comparing decision times using ORCTs**

An ORCT to address Question 2 directly would compare two ADIs that differ only in decision time. Both arms would begin with the same initial treatment, but participants assigned to one arm would be offered rescue if classified as nonresponders at week 2, and those in the other arm would be offered rescue if classified as nonresponders at week 4. The arms would then be compared in terms of their overall average long-term outcome (e.g., cannabis use reduction from baseline to week 10). As in the ORCT for comparing cutoffs described earlier, the analysis of this trial should focus on marginal means, combining responders and nonresponders within each arm.

If more than two candidate decision times are under consideration, it might be reasonable to have more than two study arms, such as weeks 2, 4, 6, and 8. Alternatively, the investigator could sequentially randomize participants as shown in Figure 1a. First, at week 2, all participants are randomized to (a) "Decide Now" or (b) "Decide Later." At weeks 4 and 6, those previously assigned to "Decide Later" would again be randomized to either "Decide Now" or "Decide Later." At week 8, everyone still in "Decide Later" would transition to "Decide Now." This design can be conceptualized as a SMART because participants are randomized more than once. Alternatively, the sequential randomizations could be replaced by a single four-arm initial randomization to the four decision times under consideration. In this scenario, investigators may choose either to disclose the assigned decision time immediately or temporarily conceal it, depending on their research objectives [25]

Allocation of participants among the decision times must be planned strategically to prioritize important comparisons. If each random assignment to "Decide Now" versus "Decide Later" is made with 50% probability, then about 50% of participants will have their decision time at week 2, 25% at week 4, 12.5% at week 6, and 12.5% at week 8, as in Figure 1a. This reduces power for comparisons among the later weeks. If the investigator desires balanced (25%) allocations across all four times, then each allocation must be adjusted accordingly, as in Figure 1b. In contrast, suppose the investigator is primarily interested in whether to decide at week 2 *versus later*. Then the week 2 comparison may be the highest priority, so the allocation in Figure 1a may be more efficient, maximizing power for the first comparison at the expense of later ones.

Regardless of the allocation, if the available sample is being divided among more than two arms, the sample size per arm becomes smaller. This limits power for pairwise comparisons, which might be important. A linear main effect of time (later vs. earlier) is inadequate for estimating an optimal time (which could be in between). As a compromise, a parsimonious model can pool data across arms (e.g., by assuming a quadratic relationship between decision time and expected final outcome). An optimal decision time can then be estimated from the model parameters.

**Selecting assessment times versus selecting decision times**

In the app usage example, assessment time is not a separate practical concern from decision time, because app usage can be automatically and continuously monitored. However, in other situations, the choices of when to make assessments and when to use the assessments to make rescue treatment decisions may be distinct design issues. Both must be informed by theoretical and practical expectations about the changes to be observed [26].

Additionally, there may be one or many assessment times. This involves practical tradeoffs. Each time a variable is assessed, there may be some burden or cost (perhaps small if done passively with a wearable sensor, or larger if it requires a clinician visit), but useful information could be gained. Assessment may even be a part of the therapeutic or preventive treatment. Food tracking in a weight loss study may serve both to encourage self-reflection about food intake and to provide tailoring information indicating when further assistance may be needed.

Although they are conceptually different and treated here as distinct components, assessment and decision time are closely related. Assessment time must occur before the decision time in order to provide useful information for tailoring. In turn, investigators may want to decide about rescue treatment as soon as possible after assessing need, to avoid wasted time and disengagement. As with the other questions, either the SDA or the ORCT approach could be used to compare assessment times. When performing SDA, an investigator must work with the assessment times used in the original study, but in an ORCT, assessment times can be randomized.

## Selecting Observed Variables

Besides choosing the timing and/or cutoff for a given observed variable, an investigator may have questions about which of multiple observed variables, or which combination of them, to use for tailoring. In the cannabis ADI example, one option is cannabis usage, e.g., defining nonresponders as those who are using cannabis more than once weekly on average by week 4. A second option is app usage, e.g., defining nonresponders as those who use the app less than once weekly by week 4. A third option combines both: e.g., those who used cannabis more than once weekly *and* used the app less than once weekly by week 4. These three possible tailoring variables correspond to three possible ways to operationalize nonresponse.

For now, suppose that each observed variable is already associated with a known reasonable cutoff, assessment time, and decision time, and the goal is to answer:

***Question 3*:** In an ADI starting with the app alone and then offering coaching to nonresponders at week 4, is it better (in terms of the overall average expected outcome at week 10) to define nonresponders as those who (a) use cannabis more than once weekly; (b) use the app less than once weekly; or (c) both?

As before, different approaches to addressing this question offer different strengths and limitations.

**Comparing observed variables using SDA**

Secondary data from a study involving the initial treatment only (with no rescue treatment) cannot directly answer Question 3. Such data could be used to predict who is likely to fail without rescue. However, among those who are likely to *fail without* rescue, the data would not predict who would be likely to *succeed with* rescue. In contrast, if some individuals received the rescue treatment, it may be possible to investigate whether the observed variables qualitatively interacted with it, i.e., whether they allow for prediction of whose outcome would or would not improve if offered rescue. Even then, participants might not have received the treatments in a way that is compatible with a proposed ADI, in terms of cutoff, decision time, etc.

For empirically choosing a tailoring variable, a dataset should contain some individuals who receive the proposed rescue treatment and some who do not, for low and high values of each observed variable under consideration. If rescue treatment was not randomized, it may be necessary to model and adjust for the process by which it was assigned [21, 27], involving technical considerations beyond this paper. Methods for using secondary datasets to estimate the effectiveness of decision rules are an ongoing research topic [e.g. 28]. If the secondary dataset is not adequate for an investigator's research questions, an ORCT may be appropriate.

**Comparing observed variables using ORCTs**

An ORCT for Question 3 could involve a single randomization to one of three ADIs that are similar in the initial treatment, decision time, and rescue treatment, but differ in the observed variable used to define response status, along with the appropriate cutoff for that variable (see Figure 2a). Suppose all three arms begin with app alone. Participants in the first arm are offered coaching if they use cannabis more than once weekly by week 2. Those in the second arm are offered coaching if they use the app less than once weekly by week 2. Those in the third arm are offered coaching if they use cannabis more than once weekly *and* use the app less than once weekly. The arms are then compared on their overall marginal average long-term outcome.

A challenge arises if the number of possible observed variables is too large to have one study arm for each separate variable. In this case, an alternative ORCT could begin with the same initial treatment as before. At week 2, each participant could then be randomized to one of two arms: augment treatment or do not augment (Figure 2b). Variables of interest would be measured and recorded at and before week 2 but would not affect the random assignment. The usefulness of the observed variables for tailoring could be investigated by testing whether they moderate the effect of augmented treatment. If there are several significant moderators, then one or two of them that have the strongest qualitative interactions with augmented treatment, or that seem most practical (e.g., easiest to measure in practice, or simplest to explain to patients), might be chosen. An important limitation of this trial is that participants are chosen entirely at random to be offered or not offered the augmented ("rescue") treatment, without tailoring by any measure of need; this might not be ethical or practical. Another limitation is that now tailoring variables must be chosen using moderation analyses (testing interactions with covariates), which may involve small effect sizes and limited power and precision. The main effect of the randomization in this single-randomization ORCT only tests whether, *on average*, augmenting treatment is more effective than not augmenting. If this question is not of primary interest, then a design that directly randomizes rescue versus non-rescue is less attractive.

## Answering Questions About Multiple Components

An investigator might wish to investigate two or more tailoring-related questions (e.g., choice of tailoring variable, decision time, and/or cutoff) together. These questions might be logically interrelated (e.g., choosing a meaningful cutoff requires first specifying the variable to which it applies), or the researcher may be interested in interactions (e.g., whether different observed variables are more useful at different decision times). For the ADI described above, suppose the investigator would like to answer three questions:

> Question 4: To maximize the expected outcome (cannabis use reduction by week 10): (4a): Which observed variable should be used to define nonresponse (cannabis use, app usage, or both)? (4b) What cutoffs of these variables should be used (e.g., how many app uses or cannabis uses per week)? (4c) When should a decision be made (e.g., week 2, 4, 6, or 8)?

As before, these questions could be addressed by either using SDA with strong assumptions or using an ORCT. However, answering multiple questions together will require stronger assumptions or a more complex ORCT.

### Answering multiple questions using SDA

To answer all of Questions 4a-c together, the investigator would need sufficient data to estimate the expected outcome for each ADI resulting from the $3 \times 2 \times 4 = 24$ combinations of variables, cutoffs, and decision times under consideration. It may be difficult to find a dataset in which the outcome is observed with and without rescue treatment under each combination, especially for a novel rescue treatment. Using datasets that include only some of the possible combinations would require additional assumptions (e.g., that certain interactions are negligible).

If only two (rather than three) questions are being considered, then it might be easier to find suitable data. However, if the existing dataset is insufficient to address these questions, then a separate ORCT may be warranted.

**Answering multiple questions using ORCTs**

A variety of experimental designs, each with advantages and disadvantages, are available for testing questions 4a through 4c. It is easier to begin by considering two of the questions at a time. For example, suppose that in addition to the most effective decision time (week 2 or week 4), investigators want to know the most effective cutoff (app usage $c = 1$, or $c = 2$ times per week on average). Conducting a separate ORCT for each question (e.g., first choosing a variable, then a decision time, then a cutoff, based on separate studies) would involve arbitrary decisions, because no choice can be operationalized without the others (e.g., a decision time requires a cutoff and vice versa). Separate studies would also be an inefficient use of resources and provide no information about interactions between the components. [10] A multiple-factor experimental design, such as a classic factorial trial, a SMART, or a hybrid, can be useful in studying multiple components of the tailoring variable and their interactions simultaneously.

In the ORCT shown in Figure 3a, participants are initially randomly assigned a decision time, and at the same time they are also randomly assigned a cutoff, just as in a 2×2 factorial design. Later, those who are classified as nonresponders, based on their assigned cutoff at their assigned time, are offered rescue, while responders continue with initial treatment. The effects of cutoff, decision time, and their interaction are then estimated. With a slightly different implementation for one of the randomizations (assigning the cutoff after assigning the decision time rather than simultaneously), the same design can be conceptualized as a SMART, as in Figure 3b.

Of course, if either the 3a or 3b version is to be ethical and feasible, each cutoff considered must be reasonable for each decision time. A related caveat is that the decision time determines the interval over which the tailoring variable can be assessed. This would have made interpretation more challenging if, instead of *average* app use occasions per week, the tailoring variable had been total uses over all weeks. Then "$c = 5$ uses by week 2" would be very different from "$c = 5$ uses by week 4." Rescaling cumulative variables by time (using average or rate instead of total) would likely make the main effects more interpretable, although an interaction could still occur.

It is possible to further extend the factorial design to all three questions. This might be done via a complete or fractional factorial design assigning participants to combinations of 3 factors: observed variables (3 levels), cutoffs (2 levels), and decision time (4 levels). However, statistical power may be a challenge here. Factors with more than two levels require larger sample sizes. Also, some main effects in this context might not be useful scientifically (e.g., it might not be meaningful to estimate the effect of cutoff, averaging over which variable was being cut off or when), but adequate power for pairwise comparisons among all 24 conditions is impractical.

A less direct, perhaps more feasible, approach could involve repeatedly randomizing participants to either rescue now (Rescue) or stay the course (Wait) at each of weeks 2, 4, 6, and 8, regardless of the observed variables, for all participants who have not yet been offered the rescue. Notice that the contrast is now Rescue (regardless of response status) or Wait, instead of Decide or Wait. At baseline, and immediately before each possible decision time (2, 4, 6, 8 weeks), observed variables (cannabis and app usage) would be measured, but randomization would be done independently of them. This approach could be seen as an incomplete factorial design with a randomization to rescue versus not rescue, and a randomization of decision time nested within the rescue arm only. It could alternatively be seen as a SMART in which randomization is not conditional on any observed tailoring variable except for past randomization itself (i.e., individuals previously randomized to rescue are not subject to future randomization). This design allows *post hoc* exploration of many covariates as possible tailoring variables, but it might not provide adequate power to decide among them except on an exploratory basis. A robust conceptual model is essential to guide the research questions and maximize the utility of the findings.

## Combined Questions about Tailoring Options and Treatment Options

In the above examples, the initial or rescue treatments were assumed to be chosen in advance, presumably based on prior evidence demonstrating their effectiveness. Only the cutoff was in question. However, investigators might also have questions about which candidate initial and/or rescue treatments are most effective, in addition to questions about tailoring variables.

### Answering combined questions using SDA

Questions combining tailoring variable options and treatment options can similarly be addressed in SDA using causal modeling techniques [20], with the same general approach and limitations, as for multiple questions about tailoring. As before, there may be a tradeoff between the breadth of the questions versus the requirements (either in terms of the data, or the strength of simplifying assumptions) to address them; technical details are beyond the scope of this paper.

### Answering combined questions using ORCTs.

ORCTs have already been employed to address questions about both tailoring and treatment options, i.e., *whom* to rescue and *when* or *how* to rescue them. For example, the ExTENd trial [1, 12, 29] was a SMART design in which all participants were offered the same initial treatment, and the first randomization was to high (5+ heavy drinking days) or low (2+ days) cutoffs for defining nonresponse. Nonresponders were later re-randomized to one of two types of rescue treatment. Data from this SMART can be used to compare cutoffs, compare rescue options, and investigate the interaction between cutoff and rescue options. Similarly, the BestFIT SMART randomized participants to decision times and to types of rescue [30], as did a pilot SMART on treating adolescent depression [4].

Suppose that in addition to the questions motivating the design in Figure 3, investigators wish to determine which of two rescue options is more effective for nonresponders. The design can be extended to a hybrid factorial-SMART wherein all participants are randomized to decision times and cutoffs, and nonresponders are later re-randomized to rescue options. This design can be implemented either by randomizing decision time and cutoff together at baseline (Figure 4a) or by randomizing them sequentially (Figure 4b).

More complicated designs are also possible. A hybrid factorial-SMART [16] could combine a factorial experiment at the initial randomization time (e.g., 2 initial treatments × 2 decision times, or × 2 levels of cutoff) with one or more later re-randomizations to different kinds of rescue treatment (e.g., moderate vs. intense) for individuals classified as nonresponders according to their arm's definition of nonresponse. Methodological research is ongoing on how best to design hybrid optimization trials and analyze the resulting data.

## Directions for Future Research

The preceding sections outlined various research questions about decision rules in ADIs, for deciding which participants should be offered a rescue treatment. The questions involved selecting the observed variables, times, and cutoffs to be used in distinguishing "nonresponders" (at higher need of rescue treatment) from "responders." We argued that a randomized experiment, where feasible, would offer the clearest and most interpretable information. An observational study (or SDA of a previous randomized study) may require stronger assumptions, especially if the intended rescue treatment is an intervention not available to participants in the original study.

Importantly, this manuscript emphasizes that the SMART is not the only useful design for optimizing ADIs; depending on the research questions, other multifactor designs, including classic factorial designs or hybrid factorial–SMARTs, may be appropriate. When there are multiple simultaneous questions, experimental approaches such as the hybrid factorial-SMART are available. However, there can be a tradeoff between the breadth of questions to be answered and the statistical power and precision available for answering them.

There are many open directions for methodological research on choosing the definition and timing of tailoring rules. In this paper, we only considered the development of an ADI with a single decision time. However, some ADIs are more flexible and allow rescue to be offered at any of multiple possible decision times, as soon as a participant meets criteria for nonresponse (e.g., [31, 32]). For example, participants in a cessation study might transition to being offered in-person coaching the first week after they report a lapse to substance use. Furthermore, the designs and recommendations considered in this paper have focused on ADIs involving relatively few decisions, made on a relatively slow time scale. However, many interventions are now digitally delivered and can adapt on a much faster timescale (e.g., daily, hourly). These just-in-time adaptive interventions (JITAIs) [see 15, 17, 33] necessitate further methodological research to establish guidelines for optimizing tailoring variables within rapidly fluctuating states and contexts.

Other research could focus on how to consider multiple outcomes such as costs and benefits. Possible ad-hoc approaches include weighting outcomes, penalizing outcomes to reflect costs (modeling cost-effectiveness instead of effectiveness), or constraining cost by a fixed budget. Cost is particularly relevant to cutoffs [19]. Even if rescue would be at least slightly effective for everyone, it might be infeasible to provide for everyone. The cutoff determines not only the definition of nonresponse but also the proportion of participants deemed nonresponders, so stricter constraints on cost would likely call for stricter cutoffs. A flexible Bayesian approach for estimation and decision making, Decision Analysis for Intervention Value Efficiency (DAIVE), was recently proposed for selecting components in factorial ORCTs. [34, 35] DAIVE could be adapted to questions about tailoring variables as well.

This paper has not addressed **uncertainty estimation.** We focused on trying to select one best tailoring rule (e.g., best combination of components such as decision time and cutoff), but perhaps several rules might be almost equally good. In a decision-focused analysis, it is sometimes necessary to choose a good-enough intervention, even if the precise best is inconclusive. However, conclusion-focused analyses require an uncertainty expression, such as standard errors, *p*-values, Bayes factors, confidence intervals, credible intervals, or equivalence sets [36].

A related issue is **power and sample size planning** for designs such as those described in this paper. Although some of our examples require only simple experimental designs for which well-known power and sample size formulas exist (e.g., a two-arm randomized trial), specifying a hypothesized (or minimal clinically significant) effect size [see 37] would still require an unfamiliar new perspective. When comparing possible cutoffs, the effect size of interest for comparing them is not that of the initial treatment, nor of the rescue treatment, but of the *choice of cutoff for assigning* the rescue treatment. If two decision times are being compared, then the effect size driving the sample size requirements is the effect of the timing of the decision, not the decision itself (see the Online Supplementary Material). If different arms represent different observed variables, then the effect size is that of the choice of variable on ADI effectiveness. Thus, eliciting or interpreting "effect size" could be challenging.

The more complex designs considered in this paper require consideration when planning sample size, because the power for some of the factors may depend on the levels to which other factors are set. For example, consider a SMART in which both the cutoff for nonresponse and the type of rescue treatment for nonresponders are randomized. The effective sample size (and hence power and precision) for comparisons of the relative performance of ADIs often depends on the proportion of nonresponders [38] because they are the only participants offered rescue treatments. In this example, the proportion of nonresponders could differ strongly between the two initial arms, because of the different cutoffs for determining nonresponse, and this could affect power calculations. We have assumed that the choice of cutoff is aimed at developing the most effective tailoring rule, regardless of whether this tailoring rule will be used in a future SMART or applied directly in clinical practice. However, the implications of a choice of cutoff on the ability to answer other important questions, in either the present or a future research study, must also be considered. Sample size has implications for the ability to empirically choose tailoring rules, and the tailoring rule has implications on the sample size needed for other analyses.

**Conclusion**

Investigators planning to develop an ADI often have questions about how best to construct a tailoring variable. These include which variable(s) to use, when to assess and use them, and what cutoff to use in mapping an observed variable to a treatment decision. These questions may be investigated using SDA (which may be less costly) but could be answered more directly with an ORCT. The ORCT could be a two-arm trial, a classic factorial design, a SMART, or a hybrid factorial-SMART. Using the latter designs, questions can be investigated at once, often more thoroughly and/or efficiently than using separate studies. [10, 16] Different approaches can answer different questions and require different assumptions; the first step is to clearly articulate the scientific questions.

**Table 1: Examples of Correlational versus Causal Questions about Tailoring**

| **Component** | **Example correlational question that can be answered via secondary data analysis (SDA)** | **Example causal question that can be answered with optimization randomized controlled trials (ORCTs)** |
|---|---|---|
| **Cutoff** | How does the average final outcome without rescue differ between those who accessed the app less than once weekly, compared to all who accessed the app less than twice weekly? | Should participants be offered rescue treatment if they accessed the app less than once per week, or less than twice per week, if the goal is to optimize the average final outcome? |
| **Decision Time** | How does the average final outcome without rescue differ between those who accessed the app less than twice weekly by week 4, compared to all who accessed the app less than twice weekly by week 2? | Should participants be offered rescue treatment if they accessed the app less than twice per week by week 4, or less than twice per week by week 2, to optimize the average final outcome? |
| **Observed Variable** | Which is more strongly predictive of average final outcome without rescue: marijuana use by week 4, or weekly app usage by week 4? | Should rescue be offered based on marijuana use by week 4, or weekly app usage by week 4, to maximize the average final outcome? |

**Figure 1**

**1a. Design for Primary Question about Week 2 Decision**

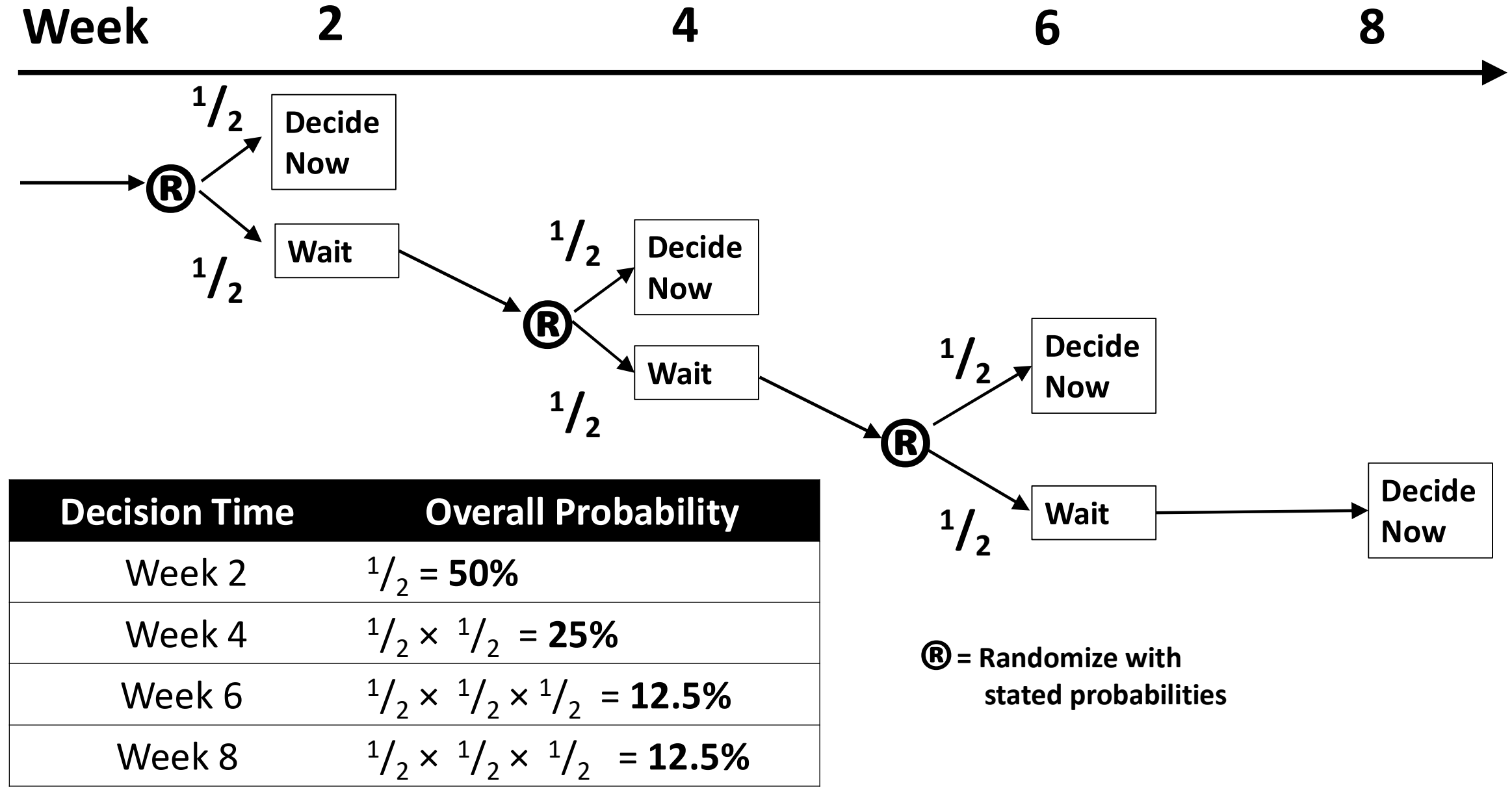

| Decision Time | Overall Probability |
|---|---|
| Week 2 | $^1/_2$ = **50%** |
| Week 4 | $^1/_2 \times ^1/_2$ = **25%** |
| Week 6 | $^1/_2 \times ^1/_2 \times ^1/_2$ = **12.5%** |
| Week 8 | $^1/_2 \times ^1/_2 \times ^1/_2$ = **12.5%** |

**1b. Design for Primary Question about Overall Decision Timing**

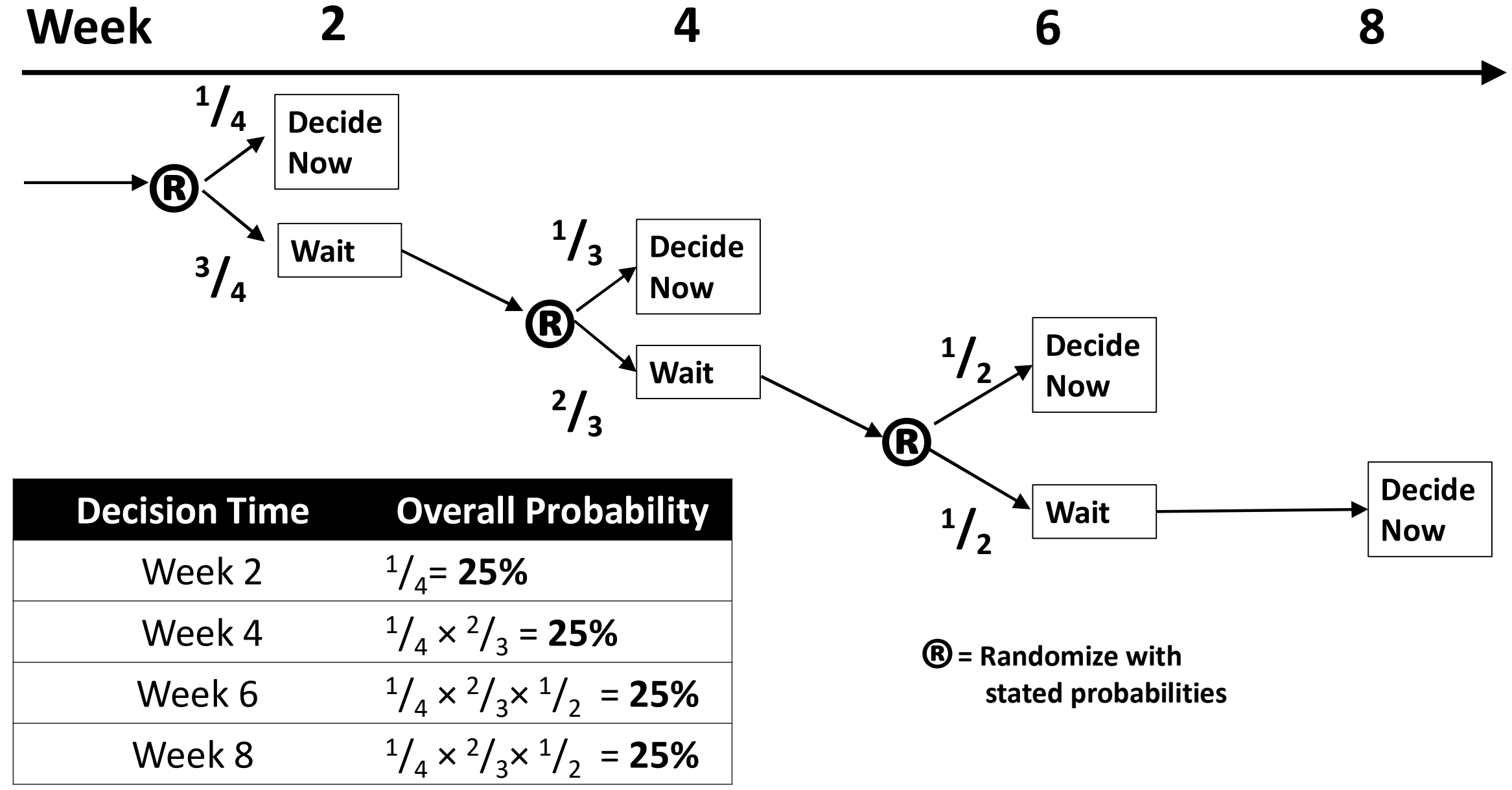

| Decision Time | Overall Probability |
|---|---|
| Week 2 | $^1/_4$= **25%** |
| Week 4 | $^1/_4 \times ^2/_3$ = **25%** |
| Week 6 | $^1/_4 \times ^2/_3 \times ^1/_2$ = **25%** |
| Week 8 | $^1/_4 \times ^2/_3 \times ^1/_2$ = **25%** |

**Figure 2**

**2a.** Multiple-arm design directly comparing candidate tailoring variables

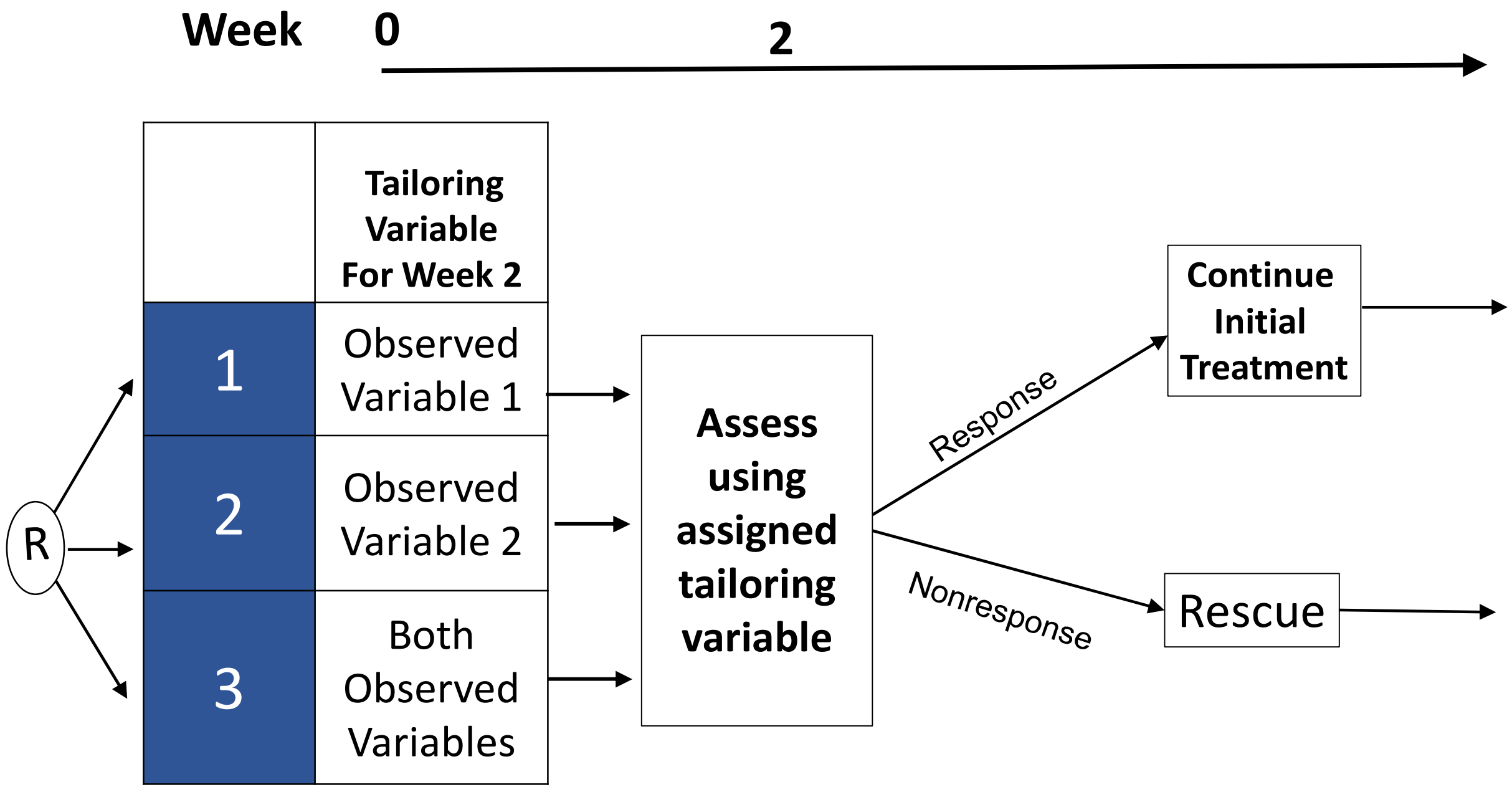

**2b. Two-arm randomized controlled trial of treatment augmentation, with assessment of moderation as an exploratory aim**

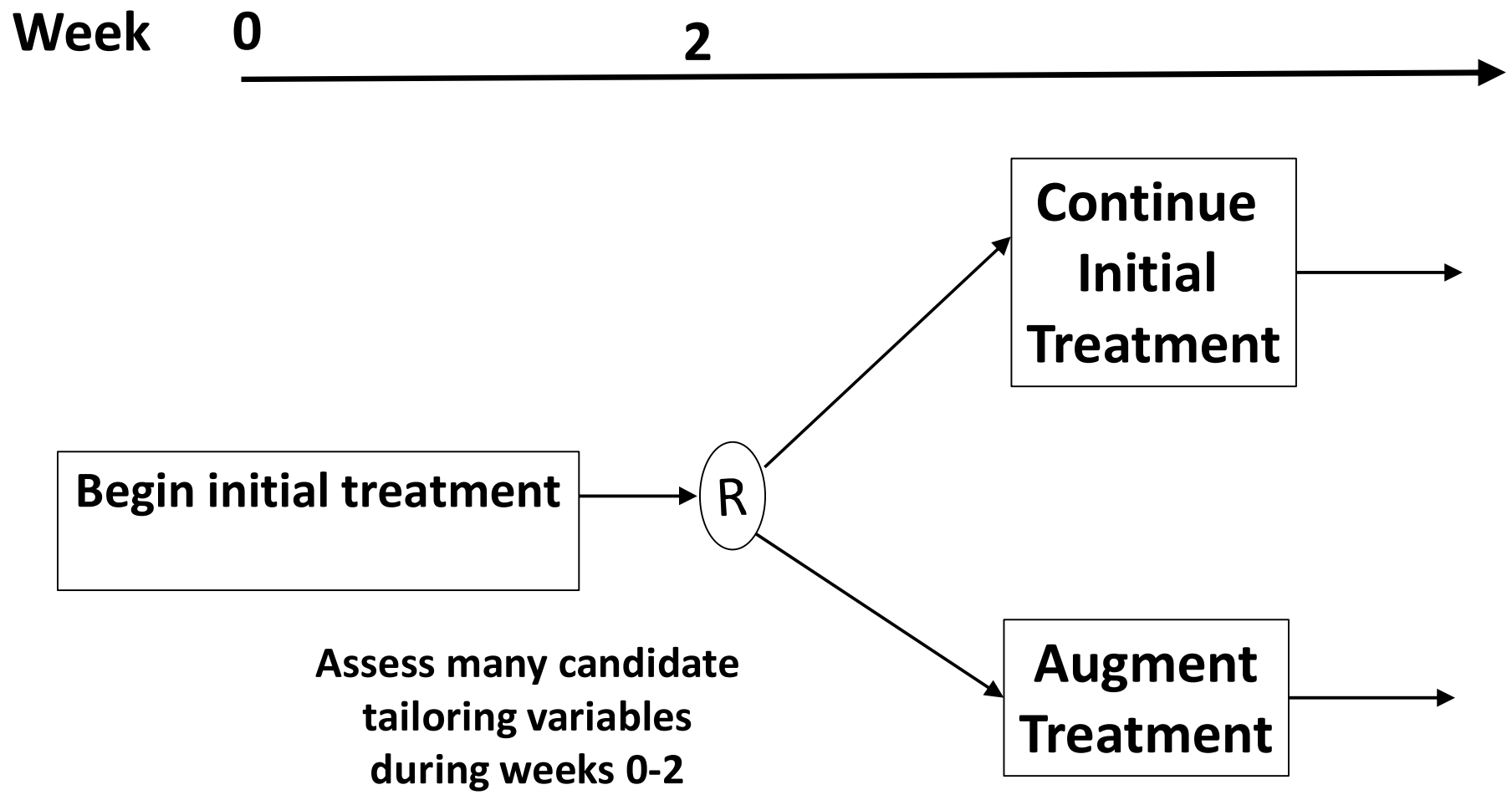

Note: All participants begin with the same initial treatment. Circled R denotes randomization.

# Figure 3

## 3a. A factorial experiment comparing decision times and cutoffs, with assignments at program entry

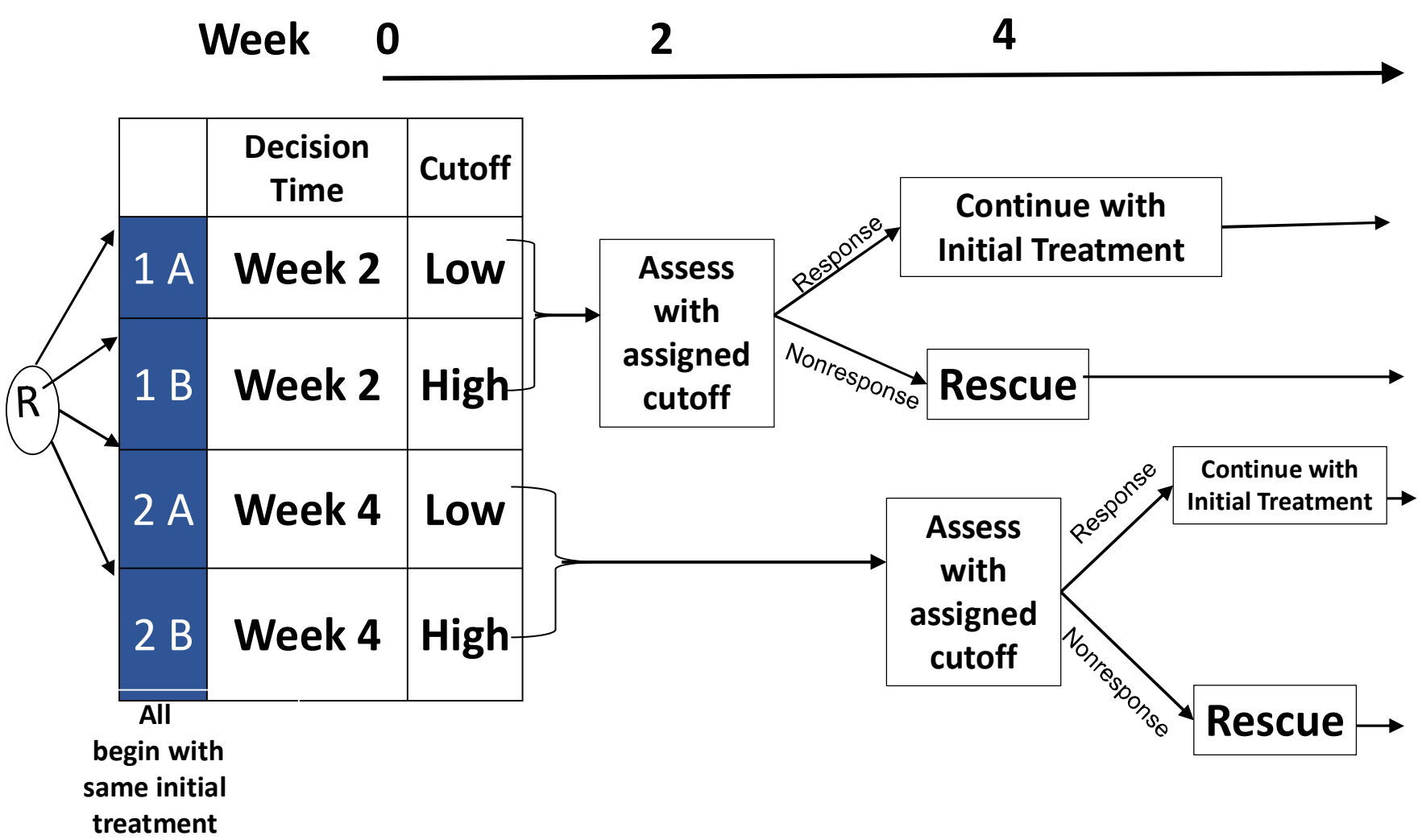

## 3b. A conceptually equivalent SMART comparing decision times and cutoffs, with assignments at or after program entry

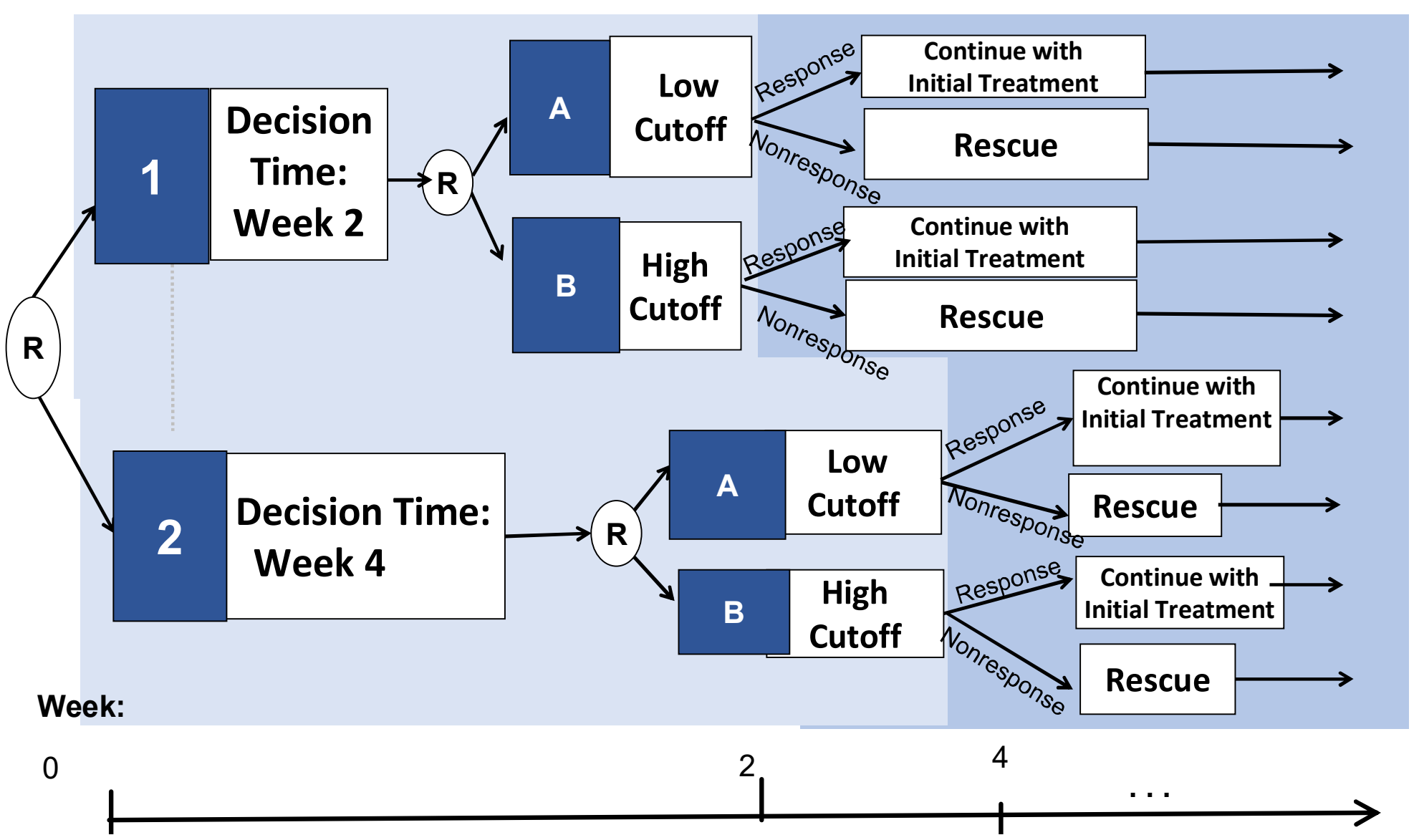

Note: Circled R denotes randomization.

## Figure 4

### 4a. Hybrid Factorial-SMART Randomizing Decision Time, Cutoff, and Rescue Type

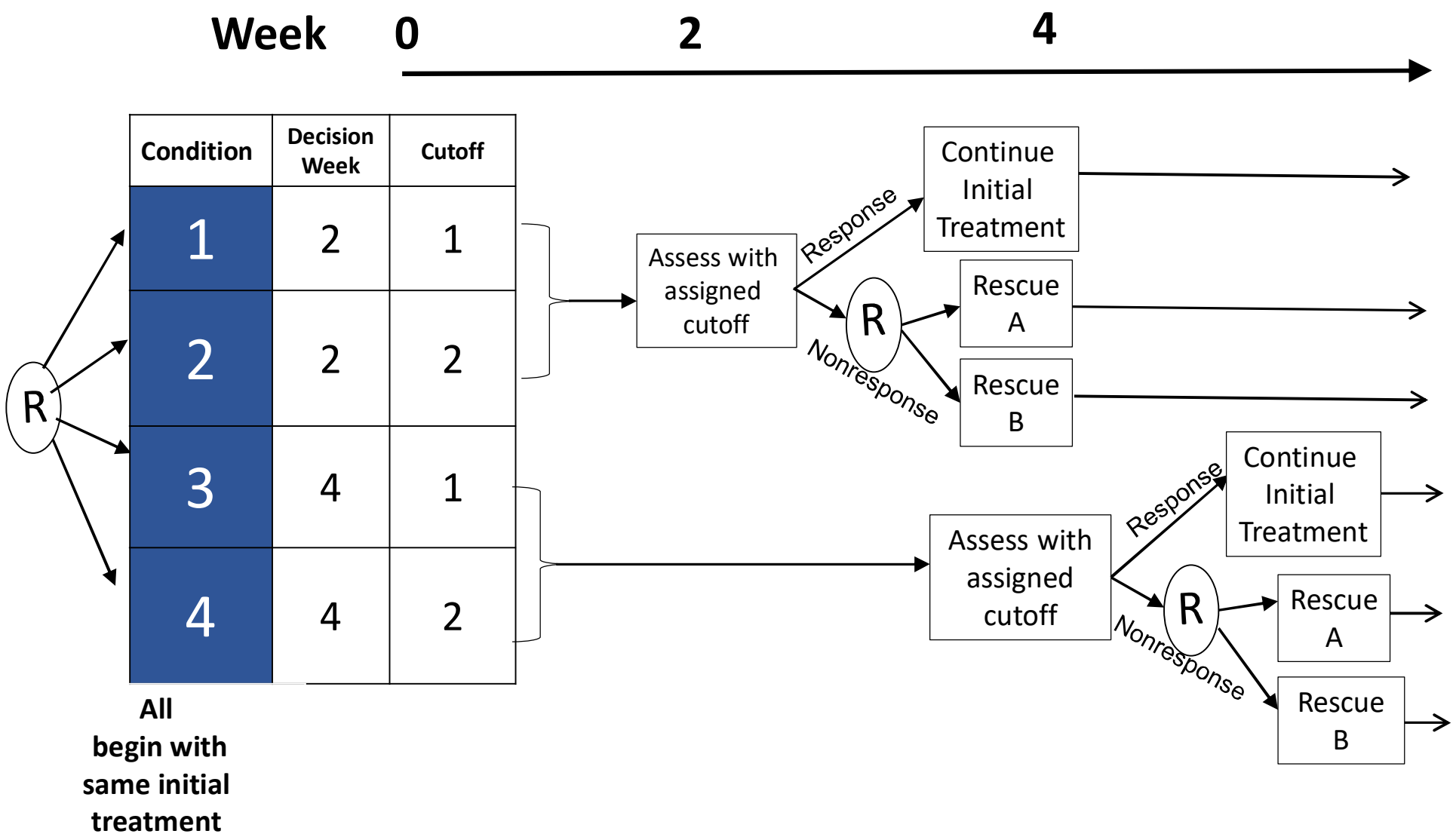

### 4b. Conceptually equivalent SMART Randomizing Decision Time, Cutoff, and Rescue Type

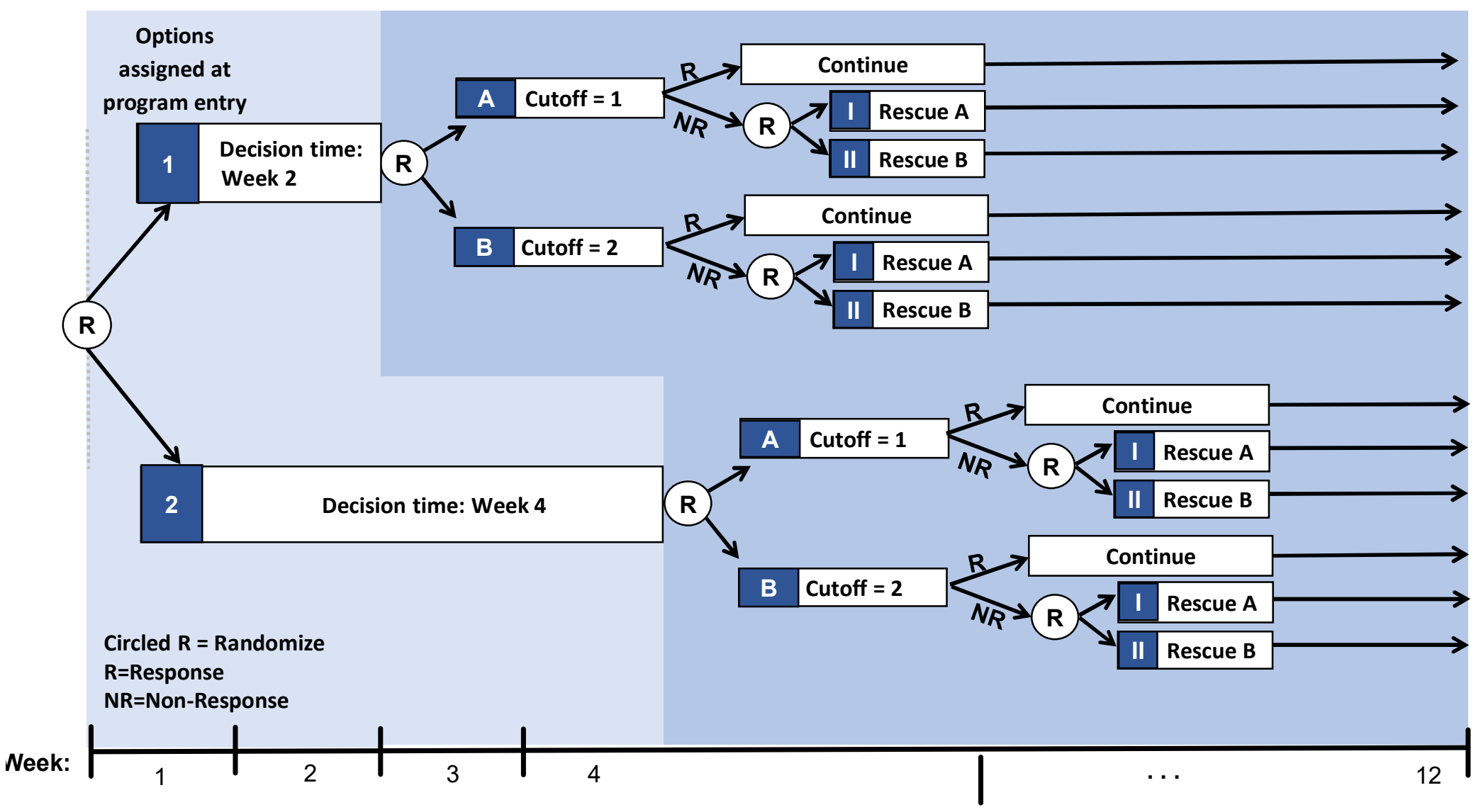

Note: Circled R denotes randomization. R and NR denote response and nonresponse.

**Online Supplementary Material for**

**"Constructing Evidence-Based Tailoring Variables for Adaptive Interventions"**

**Details on Specifying Pathways for Conceptual Models of Tailoring Variables**

In the Selecting a Cutoff section, the researcher wants to estimate the difference between $E(Y^{(1)})$ and $E(Y^{(2)})$, where $E(Y^{(c)})$ represents the expected average outcome if all individuals were treated with the adaptive intervention having cutoff $c$. It may seem unintuitive at first to consider the cutoff, rather than the rescue treatment itself, as an exposure. To illustrate, let $A_i^{(c)} = 1$ if individual $i$ will be given rescue treatment, and $A_i^{(c)} = 0$ otherwise, given that the cutoff is set to $c$. In the study under consideration, $A_i^{(c)}$ is set by design to equal $1\{O_i < c\}$, where $1\{\cdot\}$ is the indicator function (1 for true, 0 for false). This implies that changing $c$ for an adaptive intervention would affect the distribution of $A$, which in turn affects $Y$. Thus, $A$ is a mediator of $c$, as in Figure A1. In the figure, $c$ represents the choice of cutoff, $O$ represents the observed variable, $A$ represents offering vs. not offering a rescue treatment, and $Y$ represents the outcome. Dashed lines indicate a deterministic but nonlinear relationship. The error (disturbance) of Y is not shown for simplicity.

**Figure A1: Conceptual Model for Causal Effect of Cutoff on Outcome**

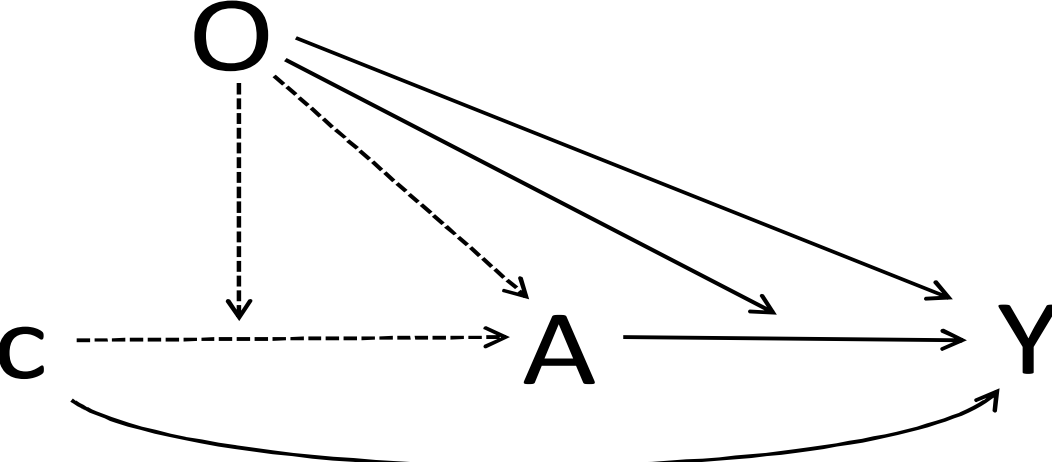

The choice of $c$ can have an effect through any of three pathways.

- **Direct effect of $c$ on $Y$**. If assignment is non-blinded, then perhaps a higher cutoff may lead either to better or worse performance for motivational reasons.
- **Indirect effect of $c$ on average $Y$ though increasing average $A$.** The cutoff impacts performance by determining the proportion of individuals being offered the rescue treatment.
- **Indirect effect of $c$ via interaction between $A$ and $O$ (moderated mediation).**. The potential effect of the rescue treatment differs for subgroups defined by $O$, in such a way that changing the cutoff changes the proportion of people receiving the level that is more effective for them.

Thus, although moderated mediation is not necessary for the cutoff to have an effect, it is the pathway which best fits the intuition of tailoring or adaptivity. There are at least three ways in which a qualitative interaction between $O$ and $A$ could occur:

- **Directly counterproductive effects**. Some aspect of rescue treatment might be beneficial for certain subgroups but iatrogenic for others.
- **Demoralization from overtreatment**. An individual might become overwhelmed, bored, or disengaged if offered unnecessary treatment.
- **Similar raw outcome but increased cost or burden**. A rescue treatment which costs time or money, but has no effect on the individual's raw outcome, would have a negative effect on a burden-adjusted or cost-adjusted outcome.

Constraints may relate to feasibility and/or ethics. Stepped care generally constrains the final intervention to give more (or at least not less) care to individuals in greater need.

In the Selecting a Decision Time section, a hypothetical investigator wished to compare $E(Y^{(2)})$ and $E(Y^{(4)})$, where now $E(Y^{(K)})$ represents the expected average outcome if all individuals were offered the adaptive intervention having decision time $K$. This is the average causal effect of the choice of decision time. As before, multiple pathways could contribute.

1. **More or less overall use of rescue treatment**. For example, suppose the definition of nonresponse in a smoking cessation study is more than $c = 2$ total lapses (uses) by the decision time. In this case, a later time gives more opportunity for lapses.
2. **Expectancy effects**. Without blinding, the awareness of an earlier or later evaluation time might shape the participant's emotions and behaviors.
3. **More effective rescue when delivered earlier**. An earlier decision point implies that rescue treatment can be delivered earlier. This might enable the rescue to have a larger effect on those who do receive it, e.g., participants are not yet demoralized.
4. **More accurate tailoring**. Later measurement provides more accurate information for identifying individuals who are more likely to benefit from the rescue (e.g., those who need it more or are more receptive) than others.

Ignoring the less intuitive pathways 1 and 2, the overall effect of $K$ may be a tradeoff between pathways 3 (quick action) and 4 (careful targeting). As a caveat, later decision time implies later rescue treatment. If the primary endpoint of the study is, say, 10 weeks from beginning of *initial* treatment, it would still be a varying number of weeks from the beginning of *rescue* treatment. Investigators who want to know whether the effect of rescue changes over time (delayed effects and/or attenuation) should consider repeated follow-ups, to avoid confounding "time since beginning" with "time until end."